\documentclass[article]{revtex4-1}
\usepackage{graphicx}
\usepackage{dcolumn}
\usepackage{bm}
\usepackage{hyperref}
\usepackage{amsmath,amssymb,amsfonts,graphicx,float}

\newcommand{\beq}{\begin{eqnarray}}
\newcommand{\eeq}{\end{eqnarray}}

\begin{document}

\title{Fractionalize This}

\author{Philip Phillips}

\affiliation{Department of Physics,
University of Illinois
1110 W. Green Street, Urbana, IL 61801, U.S.A.}

\date{\today}

\begin{abstract}
Precisely what are the  electrons in a high-temperature superconductor doing before they superconduct?  Strong electronic correlations may give rise to composite rather than fractionalized excitations, as is typical in other strongly coupled systems such as quark matter.
\end{abstract}

\pacs{}
\keywords{}
\maketitle

\begin{widetext}
\begin{figure}
\centering
\includegraphics[width=15.0cm]{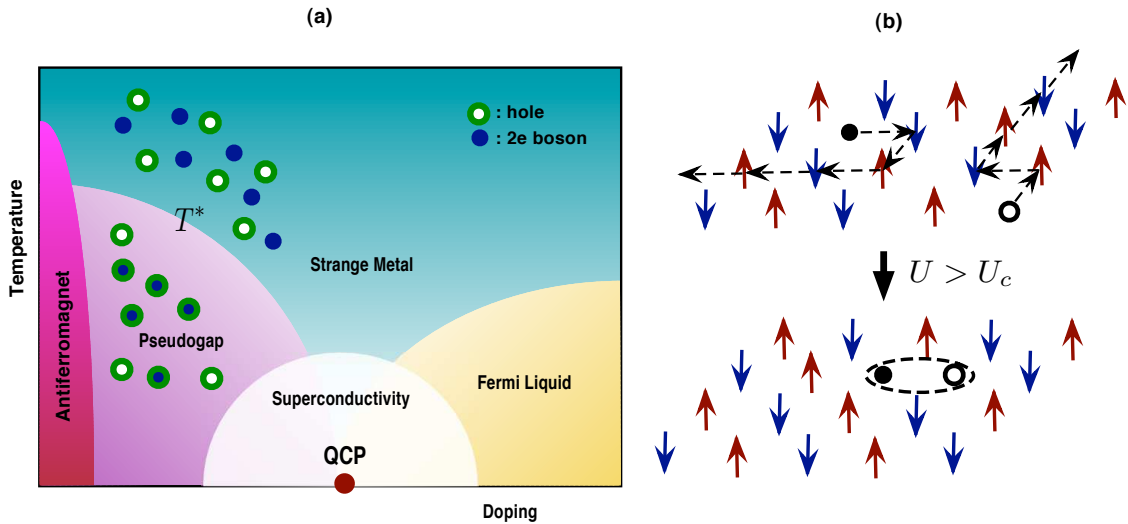}
\caption{Phase diagram and anomalous transport in the cuprate
  high-temperature superconductors.  (a) Heuristic phase diagram as a
  function of holes doped into the copper-oxide plane.  The pseudogap
  and strange metal are characterized by a depletion of the density of
states and a $T$-linear resistivity, respectively. The pseudogap terminates at a zero-temperature critical point or quantum critical point (QCP). To the right is a
Fermi liquid where weak-coupling accounts become valid. The picture
advanced here for the pseudogap to strange-metal dichotomy is an
unbinding transition involving charge 2e bosons and holes.  The charge
$2e$ boson is a collective excitation arising from the inseparability of
the high and low energy scales in the cuprates. Above the pseudogap
line, $T^\ast$, the charge $2e$ boson and a hole unbind.
Scattering of bosons and electrons above the temperature to create the
boson naturally yields $T-$linear resistivity.  (b) The
mechanism for the insulating state in the parent material, a half-filled band, as proposed by Mott. Up (down)
arrows indicate spin up (down) electrons. 
Below a critical value of the on-site interaction $U$, doubly occupied
(solid circles) sites and holes (empty circles) are free to transport.
Above a critical value of $U$, they are bound in localized pairs,
thereby preventing conduction.}
\label{tlin}
\end{figure}
\end{widetext}

24 years after the discovery of superconductivity in the copper-oxide
ceramics (hereafter cuprates), the central problem remains the anomalous properties of the
normal state.   The key anomaly is the strange metal (see Fig. (\ref{tlin})) in which the 
resistivity scales as a linear function of temperature rather than the
characteristic $T^2$
dependence of Landau's standard theory of metals. The fact that this transport anomaly persists to unusually high temperatures,
roughly 1000K, indicates  that it is a robust feature rather than some
incipient nuisance that can just be dabbed away.  As the phase diagram (see Fig. (1))
of the cuprates lays plain, the correct theory of the superconducting
state should  give rise to the panoply of phases
that emerge at higher temperatures.  That is, theories which describe
only the superconducting state or just the pseudogap, a state with
vanishing superconducting order but a gap in the single-particle
spectrum  nonetheless, are clearly inadequate.  This suggests that the standard
guiding principle of model
building in which only $T=0$ states are relevant fails in this problem because neither
the pseudogap nor the strange metal appear necessarily as zero-temperature phases
in the cuprate phase diagram (at zero magnetic field).   Instead,
these phases emerge from the
high-temperature correlated electron liquid, or charge vacuum, of the cuprates. 

{\bf Non-Fermi Liquid lost then found}

Whatever this charge vacuum is, we know it must be a non-Fermi liquid.  In modern terms, Landau's theory of a Fermi
liquid rests on the key fact that all renormalizations arising from
screened Coulomb interactions preserve the Fermi surface. Hence,
as long as a Fermi surface exists and electrons are the charge
carriers, the strength of the short-range repulsions is irrelevant.
Consequently, any theory that claims $T-$linear resistivity from a
perturbative analysis around a Fermi surface is moot. As a result, it
is unfortunate that much of the current activity in the field of
high-temperature superconductivity has focused on reclaiming a Fermi
liquid picture in the region of the phase diagram precisely where it
is least likely to apply.  The primary impetus for this is the observation of
coherent quantum oscillations indicative of closed orbits in
high-magnetic field experiments\cite{qoscill} on a class of cuprates.  Indeed, such experiments
are central to the story at high magnetic fields.  But that perhaps is it.

In this regard, the recent experiment by Fournier and co-workers\cite{dama}
is important because it shows precisely how and where in the phase
diagram the Fermi liquid
picture fails. Any many-body problem is solved once the
propagating degrees of freedom are isolated.  These are the
excitations which give rise to poles in the corresponding
single-particle propagator, thereby describing coherent excitations.  Fournier and colleagues\cite{dama} used angle-resolved
photoemission spectroscopy (ARPES) to measure the strength of the pole in the electron
propagator in YBa$_2$Cu$_3$O$_{6+x}$ with truly unprecedented accuracy and found it to vanish below a
critical doping level of $x\approx 0.15$,
indicating that electrons are not the propagating degrees of freedom
at any momenta
in the underdoped region. 
While this result is not particularly surprising given the series of
experiments\cite{campuzano,zxshen} that preceeded it, it comes at a crucial time when attention
 needs to be refocused on the strong coupling physics that is the high
$T_c$ problem. 

{\bf Limits of Quantum Criticality}

Early in the theory of the cuprates, explanations\cite{varma,tallon} based on some type of
quantum criticality, that is, a phase transition at $T=0$ driven by
quantum not classical fluctuations, were invoked to explain $T-$linear
resistivity.  The principle underlying such explanations is
universality.  Namely, at the quantum critical coupling, the only
energy scale governing collisions between quasiparticle
excitations of the order parameter is $k_BT$, where $k_B$ is
Boltzmann's constant.  Consequently, the scattering  rate between electrons
scales as
\beq
\frac{1}{\tau_{\rm tr}}\approx \frac{k_BT}{\hbar},
\eeq
with $\hbar=h/2\pi$, $h$ Planck's constant.  Despite the fact that the
momentum scales contributing to the scattering rate and the
resistivity differ, a similar $T-$linear scaling was
reasoned to hold for the resistivity as well\cite{marginal}.  But is this true?
Quantum critical scenarios come in one of two types. Either the
charge carriers are the critical modes or they are not 
but instead are coupled to a degree of freedom that is\cite{hertz}.  In
the latter, it is well known\cite{hertz} that the resultant resistivity scales as
$T^a$ where $a$ is positive typically exceeding unity.
What about theories in which the charges are critical?  A few years
ago, I showed with Claudio Chamon\cite{chamon} that under three general
assumptions, 1) the charges are critical, 2) charge is conserved, and
3) a single parameter governs the scaling of all energies and
length scales, the dc resistivity must
scale as
\beq\label{scal}
\rho(T)\propto T^{(2-d)/z},
\eeq
regardless of the underlying critical theory and the statistics of the
particles. Here $z$ is a number which measures how time and spatial
increments are related. For example, if they are on the same footing,
$z=1$.  For a three-dimensional ($d=3$) system such as the
cuprates, $T-$linear resistivity obtains only when $z=-1$. This is an impossiblity as
it implies superluminal transport, thereby violating causality.
Consequently, if the charges are critical, then either charge
conservation must be abandoned, highly unlikely, or some new length
scale (new degree of freedom) must emerge that is not governed by the same dynamics that
underlies the divergence of the correlation length.  Interestingly, breaking Fermi
liquid theory in $d=2$ requires new degrees of freedom as well because simply
cranking up the short-range interactions cannot destroy the Fermi
surface.  

{\bf Mottness: Emergence of new degrees of freedom}

Consequently, the central questions are what are the new degrees of freedom, how do they arise from the
underlying microscopic theory of the cuprates, and how do they mediate
$T-$linear resistivity?  There is a highly successful account of $T-$
linear resistivity known as marginal\cite{marginal}
Fermi liquid theory (MFLT).  This approach is based on a
phenomenological proposal for
the self interaction of the electrons which explains 1) the broad spectral features seen in
ARPES experiments\cite{campuzano,zxshen} at
optimal doping and 2) also $T-$linear resistivity.   While there have been
recent attempts\cite{varma} to derive the key posit of MFLT from
microscopics, no derivation exists that captures the strong
interactions of the insulating state of the parent cuprates. This
failure has caught the attention of the string theorists who have\cite{mcgreevy}
brought to bear the gauge/gravity duality, which weds the geometry of space-time with
quantum mechanics, on this problem.  However, even this derivation\cite{mcgreevy} is phenomenological
because the desired result is obtained by tuning the scaling dimension of a set of
operators.  It is the hope that in the quantum theory, the operators
get quantized and hence acquire a definite scaling dimension.
However, because the
underlying quantum theory is not known, the precise physical content
of the tuning parameter which gives rise to MFLT 
remains an open problem.   Nonetheless, this approach demonstrates beautifully
 the inherent disconnect between a high-energy (UV) and a low-energy (IR)
description of a strongly coupled system.   To illustrate, the key claim of
 gauge/gravity duality is that some interacting quantum theories at
strong coupling in d-space-time dimensions are dual to gravity
theories in a $d + 1$ dimensional space-time with a constant
curvature, in this case one that is asymptotically anti-de Sitter
(AdS$_{d+1}$).  In spite of the fact that the quantum theory resides at the
boundary of the space-time, the $UV$ scale, all the information
regarding what the charges are doing at low frequencies is determined entirely by the near horizon metric AdS$_2
\times\mathbb R^2$ ( a space which is hyperbolic in two dimensions but flat in the other two), which is as far away from the boundary as
possible. Hence, the degrees of freedom which
emerge in the IR and govern the analytic structure of the theory have
no correspondence with those in the original charged AdS$_4$
UV limit. The same should also be true of the underlying microscopic quantum
theory which ultimately describes marginal Fermi liquid theory. This
suggests that extracting marginal Fermi liquid behaviour from the
basic model of the cuprates  might be tricky because the natural
variables which would expose this behaviour are not the bare
electrons at the UV.  In fact, the operators whose scaling dimension  gives rise
to MFLT in the gauge/gravity theory have nothing to do with
electrons.  This conundrum has motivated any number of theories in
which the electron is assumed to fractionalise.  As is the case in other strongly coupled
problems, for example mesons (bound quark
states) in quantum chromodynamics (QCD), the more likely outcome in the cuprate problem is that the excitations which emerge as being coherent are composite
not fractionalised excitations. 

Precisely what makes the charge vacuum in the cuprates unique was first
uncloaked in a series of x-ray experiments by Chen, et al.\cite{chen}
designed to measure the available phase space for adding a particle at
low energies.
They concluded that the available phase space is much larger than that expected for a Fermi
liquid. Hence, it is important to understand this effect because it
points to a clear source of the extra degrees of freedom.  In a Fermi liquid, the number of available states for
particle addition at low energy equals the number of holes
created, $x$. Since each hole represents a quasiparticle, the quasiparticle weight scales as $x$. However,
in the cuprates, the x-ray intensity at low energies increases faster than $2x$\cite{chen}.  The basics of this
effect are clear\cite{sawatzky,phillips} and intrinsically tied to the
strong interactions in the parent state of the cuprates. Undoped the
cuprates are essentially Mott insulators in that they possess a
half-filled band
but 
insulate nonetheless.  They
also order antiferromagnetically.  However, that the x-ray intensity
measured by Chen, et al.\cite{chen} increases faster than $2x$\cite{chen} follows solely from the strong
interactions rather than from antiferromagnetic order.  All such
physics which is tethered to the strong interactions and not order is
termed Mottness. Consider a simple model (due to Hubbard) for a Mott insulator in
which electrons hop on a lattice with a matrix element of magnitude $t$ but pay an energy cost $U$ when they
doubly occupy the same site.  To understand the experiments of Chen,
et al.\cite{chen}, we need to consider how the spectrum rearranges
upon the creation or removal of electrons, that is, hole doping. A
hole leaves behind an empty site. Each such empty site can be occupied
by either a spin-up or a spin-down electron. Hence, just from doping,
the empty part of the spectrum at low energies has an intensity of
$2x$.  Further, since hole doping
annihilates one state in the filled part of the spectrum, there are
$1-x$ electron states (per site) remaining
below the chemical potential.   This gives rise to a total intensity
of the lower band of $1+x$. Based on this counting ( and the Fermi
liquid precident that the quasiparticle weight should scale with the
empty part of the spectrum), it is tempting to infer that the
quasiparticle weight should scale as $2x$ divided by the total weight
of the low-energy band ($Z=2x/(1+x)$) as has been
proposed\cite{anderson}.  However, the Fournier, et al.\cite{dama} experiment
shows that this scaling fails below $x\approx 0.23$ with a vanishing
weight at $\approx 0.15$.    

How can the vanishing of $Z$ be explained? Does the Chen, et
al.\cite{chen} experiment help?
Indeed it does because we have not accounted for the total intensity
of the lower band by just counting electron states.  The key point is that not all empty sites in the
Hubbard model are created equally. Empty
sites can be created simply by electron hopping. Consider an up-spin electron jumping to a site
occupied by an electron with the opposite spin. Such empty sites automatically generate double
occupancy and even exist in the insulating
state as illustrated in Fig. (\ref{tlin}b).  The empty site left behind affects the
excitation spectrum at all energies in that not only does the number
of ways of creating doubly occupied sites decrease but so does the
effective number of singly occupied sites\cite{sawatzky,chempot,harrislange}. In addition, the number of
ways of adding a particle at low energies increases\cite{chempot}.  It has been known since 1967\cite{harrislange}
that the total weight or intensity of the low-energy band, as a result of
 mixing with doubly occupied sites (termed dynamical spectral weight transfer), {\bf increases} from $1+x$ (zero
hopping limit) to
$1+x+\alpha$, where $\alpha$ is a positive correction.
The leading term in $\alpha$ scales as $t/U$ and as a consequence is generally
ignored at strong coupling.   However, retaining it leads to an
important effect. ( Recall even antiferromagnetism is a $t/U$ effect
which of course no one ignores.)   While the intensity of the
low-energy band increases when the hopping is turned on (or $U$ is decreased), the number of
ways electrons can be assigned still remains fixed at $1+x$. Hence, dynamical spectral weight transfer poses a
distinct problem:  the
total intensity of the low-energy band exceeds the number of electrons
that can be assigned to this band. Such a mismatch cannot obtain in a
Fermi liquid or band insulator because in such systems the spectral
intensity of the bands is independent of the particle density.  As a consequence, dynamical spectral weight transfer
requires additional degrees of freedom that are not exhausted by counting
electrons alone.   Although not stated explicitly, this must have been
`known' to Harris and Lange\cite{harrislange} in 1967 because they refrained from dividing the spectrum into a
filled and empty part based on the bare electron charge: $1-x$
filled and $2x+\alpha>2x$ empty states.  If the total weight of a band
exceeds
the electron count, then what the electrons are doing is
irrelevant!  Likewise, if the phase space for adding a particle
at low energies
 exceeds the number of
ways electrons can be added to the band, then
there are ways of adding a coherent excitation at low energies that are orthogonal
to the addition of an electron.  Hence, dynamical spectral weight
transfer is directly linked to the vanishing of $Z$, thereby making it
 a property that affects ground and excited states alike.  Since the intensity exceeding $1+x$ arises entirely in strong
coupling, it should vanish once the doping level is such that
weakly interacting physics obtains.  This is the collapse of Mottness
and is signaled by a decoupling of the upper and low-energy bands
in a Mott insulator.  Some experimental evidence for this has recently
been seen by Lin, et al.\cite{chen2} and in simulations by Jarrell
and collaborators\cite{jarrell}.  

{\bf Doublon-holon unbinding: Strange Metal}

The relevant question then is: how can dynamical spectral weight
transfer and the extra degree of freedom that describes it be captured by a low-energy theory of a doped mott
insulator? The procedure put forth by Kenneth Wilson\cite{wilson} for constructing a low-energy
theory is to integrate out the unwanted high-energy stuff.  In the context of the Hubbard model, it is
typical to construct a low-energy
theory by removing double occupacy.  But as we have seen, such a
procedure does more than Wilson would advise because without double occupancy, the
dynamical correction vanishes.   While it is possible in
projected schemes to retain the dynamical correction, isolating the
dynamics it mediates is difficult because it is buried in cumbersome
operator transforms\cite{tremblay}.  Is it possible to represent the dynamical mixing
with a new collective mode? Indeed it is. This new collective mode should have
charge $2e$ since it represents the mixing with double occupancy.    
The simplest procedure is to represent the physics of the upper band by
an appropriately chosen coordinate rather than by particular electronic
configurations.  The subsequent 
Lagrangian\cite{phillips} is
quadratic in the coordinate for the high-energy scale and
hence can be integrated out exactly to yield the low-energy physics.
What results\cite{phillips} is a theory with electrons and a new collective mode, a
charge $2e$ boson. The charge $2e$ boson enters the theory initially as a
Lagrange multiplier and hence is undamped.
The low-energy physics it mediates typifies that of strong
coupling.  New charge $e$ states ( in addition to the standard
projected electron states in the lower band that give rise to the $2x$
sum rule) emerge that have internal
degrees of freedom and hence are orthogonal to a bare electron.  The
new states correspond to a bound state of the charge $2e$ boson and a
hole.   At half-filling\cite{phillips}, the bound state (depicted in Fig. (\ref{tlin})) generates a charge gap and
represents doublon-holon binding to which many previously have
attributed (without proof) the Mott
gap\cite{mott,kohn,fulde}. At finite doping, such binding persists
yielding a pseudogap. Supporting
this picture are recent oxygen K-edge experiments\cite{chen2} in which
the sign and magnitude of the temperature dependence of the dynamical
contribution to the spectral weight across the pseudogap line is in excellent agreement with the
prediction from the charge $2e$ theory .  The breakup
of the boson-holon bound state generates the strange metal (see Fig. (\ref{tlin})).  The mechanism for $T-$linear resistivity is simple within
this model. Once the binding energy of the boson vanishes, bosons are
free to scatter off the electrons. The absence of a kinetic energy
term for the bosons implies that their dynamics are classical.  The resistivity of electrons
scattering off classical bosons is well-known to scale
 linearly with temperature above the
energy to create the boson as depicted in Fig. (\ref{tlin}).  Hence, this mechanism is robust and should
persist to high temperatures. Further, the presence of two distinct
charge $e$ states at low energy, one of which generates activated
transport, naturally yields a two-fluid model which has been shown to
underlie the temperature dependence of the Hall coefficient\cite{Hall}.  Consequently, the charge 2e boson
reduction of the Hubbard model offers a resolution of the pseudogap,
the vanishing of the quasiparticle weight in the underdoped regime and the transition to the
strange metal regime of the cuprates. The precise details of the
unbinding transition (which preliminary work\cite{elp} indicates might be amenable to the gauge/gravity duality) and the role of the composite excitations in the
superconducting state remain open.

\acknowledgements This work is partially funded by NSF under Grant No. DMR-0940992. The author acknowledges support from the Center for Emergent Superconductivity, an Energy Frontier Research Center funded by the U.S. Department of Energy, Office of Science, Office of Basic Energy Sciences under Award Number DE-AC0298CH1088.

\end{document}